\documentclass[traditabstract]{aachanged}
\usepackage[english]{babel}
\usepackage{amsmath}
\usepackage{amssymb}
\usepackage{euscript}
\usepackage{cyr}
\usepackage{epsfig}
\begin{document}

\title{Influence of Gamma-Ray Emission on the Isotopic Composition
of Clouds in the Interstellar Medium}
\titlerunning{Influence of Gamma-Ray Emission}
\author{V.V. Klimenko$^{1,2}$\thanks{E-mail: slava.klimenko@gmail.com}, A.V. Ivanchik$^{1,2}$\thanks{E-mail: iav@astro.ioffe.ru}, D.A. Varshalovich$^{1,2}$\thanks{E-mail: varsh@astro.ioffe.ru} and A.G. Pavlov$^2$}
%\author{V.V.~Klimenko$^{1,2}$,
%A.V.~Ivanchik$^{1,2}$, D.A.~Varshalovich$^{1,2}$, and A.G. Pavlov$^{2}$}
\authorrunning{Klimenko et al.}
\date{Received 19 December, 2011}
%\vspace{4pt}\\

\institute{\it{$^{1}$Ioffe Physical–Technical Institute, ul. Politekhnicheskaya 26, St. Petersburg, 194021 Russia} \\
\it{$^{2}$St. Petersburg State Polytechnical University, ul. Politekhnicheskaya 29, St. Petersburg,
Russia}}

%\pagerange{\pageref{firstpage}--\pageref{lastpage}} \pubyear{2009}

%\label{firstpage}

%\begin{abstract}
\abstract{We investigate one mechanism of the change in the isotopic composition of cosmologically
distant clouds of interstellar gas whose matter was subjected only slightly to star formation processes.
According to the standard cosmological model, the isotopic composition of the gas in such clouds was
formed at the epoch of Big Bang nucleosynthesis and is determined only by the baryon density in the
Universe. The dispersion in the available cloud composition observations exceeds the errors of individual
measurements. This may indicate that there are mechanisms of the change in the composition of matter
in the Universe after the completion of Big Bang nucleosynthesis. We have calculated the destruction and
production rates of light isotopes (D, $^3$He, $^4$He) under the influence of photonuclear reactions triggered
by the gamma-ray emission from active galactic nuclei (AGNs). We investigate the destruction and
production of light elements depending on the spectral characteristics of the gamma-ray emission. We
show that in comparison with previous works, taking into account the influence of spectral hardness on the
photonuclear reaction rates can increase the characteristic radii of influence of the gamma-ray emission
from AGNs by a factor of 2–-8. The high gamma-ray luminosities of AGNs observed in recent years increase
the previous estimates of the characteristic radii by two orders of magnitude. This may suggest that the
influence of the emission from AGNs on the change in the composition of the medium in the immediate
neighborhood (the host galaxy) has been underestimated.}


\maketitle

\section{Introduction}
\label{introduction}
\noindent
The relative abundance of the light elements (H, D, $^3$He, $^4$He, $^6$Li,
$^7$Li) produced at the epoch of Big Bang nucleosynthesis is one of the criteria for
testing the standard cosmological model. According
to this model, Big Bang nucleosynthesis occurred in
the early Universe: at times from several minutes,
when the Universe was extremely hot (T$>10^9\,$K),
to half an hour, when the temperature and density
of the medium dropped to values at which the nuclear
reactions no longer proceeded. Subsequently,
the composition of the interstellar and intergalactic
medium changed mainly under the influence of astration
and radioactive decay. Therefore, by studying the
isotopic composition of distant clouds of interstellar
gas $\mbox{($z\sim2\div3$)}$, where the star formation processes
had not yet strongly affected the relative elemental
abundances, their abundances at the completion time
of Big Bang nucleosynthesis can be determined

One of the most important aspects of such an
analysis is the abundance of deuterium. In comparison
with other elements produced at the epoch of Big
Bang nucleosynthesis, deuterium has the simplest
nuclear structure and is most sensitive to a change
in the baryon density $\Omega_b$ -- one of the key parameters
of the standard cosmologicalmodel. The D/H ratio in
the interstellar medium of galaxies can be calculated
from the absorption by neutral gas of light emitted by
distant quasars (QSO) by comparing the intensities
of hydrogen absorption lines with those of deuterium
ones. When the primordial D/H ratio is estimated,
deuterium is commonly assumed to be only destroyed
in the course of stellar evolution. Therefore, any D/H
estimate is considered only as a lower limit for the
primordial D/H ratio.

Some of the first deuterium abundance measurements
in clouds of interstellar gas at high redshifts
gave large D/H ratios, $\sim10^{-4}$ (Webb et al. 1997;
Rugers and Hogan 1996), which was inconsistent
with the predictions of the standard model. An
overview of the possible mechanisms of the change
in the abundances of light elements in clouds of
interstellar gas relative to their primordial values was
provided by Gnedin and Ostriker (1992). However,
the new deuterium abundance observations (Burles
and Tytler 1998) disproving the previous ones were
in agreement with the theory, and the question about
the change in abundance was essentially lifted. The
currently available light-element abundance data
taken from Steigman (2007) are presented in Table~\ref{table1}.
Eleven absorption systems of atomic deuterium and
hydrogen have been detected in the last ten years, and
the D/H ratio has been measured in them with a high
accuracy. The derived values are consistent, in order
of magnitude, with the conclusions of the standard
model, but there are some problems. At present, only
seven systems in which the deuterium abundance
can be determined with good confidence from an
analysis of absorption lines (Pettini et al. 2008,
D/H=2.82$\pm0.20\times10^{-5}$) are used to estimate the
D/H ratio. However, even for this sample the $1\sigma$
scatter of D/H values was about 20\%, which is
considerably higher than the observational errors of
individual measurements.

It is believed that the scatter of deuterium abundances
in such clouds can be provided by both star
formation processes and deuterium depletion onto
dust grains (Linsky et al. 2006). The interstellar gas
that was subjected only slightly to chemical enrichment,
as is suggested by the low relative abundance of
heavy elements, O/H~$<$~1/10 of the value measured
in the Solar system, was studied in these seven systems.
In the model of galactic chemical evolution,
the decrease in the D/H ratio for such clouds is negligible,
which was theoretically shown by Akerman
et al. (2005), Prantzos and Ishimaru (2001), and
Romano et al. (2006). Therefore, the large dispersion
in the experimental data is most likely explained by
an underestimation of the statistical and systematic
errors (Steigman 2007; Pettini et al. 2008). Apart
from the standard theory of Big Bang nucleosynthesis,
the dispersion in the light-element abundance
observations is associated with the evaporation of
primordial black holes or the decay of long-lived X-particles.
The high-energy hadrons and photons
being produced during the decay of X-particles destroy
the light elements synthesized in Big Bang
nucleosynthesis. Effective $^4$He destruction and D,
T, $^3$He production can occur as a consequence of
this process. These processes were investigated by
Zel’dovich et al. (1977), Kawasaki et al. (2005), and
Carr et al. (2010). Stringent constraints were placed
on the primordial abundance of the long-lived exotic
X-particles.

In this paper, we investigate one mechanism of
the change in the relative composition of the medium
that was previously considered by Boyd et al. (1989)
and Gnedin and Ostriker (1992). Hard X-ray and
gamma-ray emission is present in the spectrum of
active galactic nuclei (AGNs). Propagating in the
Universe, the high-energy gamma emission interact
with the interstellar and intergalactic matter. The
photodisintegration of 4He with the production of D
can be compared with the photodisintegration of D.
This can cause both a decrease and an increase in
the relative abundance of deuterium in some regions
of interstellar gas. The magnitude of this effect depends
on the initial composition of the medium (the
abundance of $^4$He in a cloud is greater than that of D
by three orders of magnitude), the spectral hardness
of the gamma-ray emission characterized by the parameter
$\Gamma$ (see Eq. \ref{eq3}), and its flux.

\begin{table}
\caption{ Primordial isotopic composition of matter in the Universe}
\label{table1}
\vspace{2mm}
\begin{tabular}{l|c}
 \hline\hline
{Element abundance }&{Primordial value$^b$}\\
{relative to hydrogen$^a$} & {}\\
\hline\hline
~~~~~~$X_{\mbox{D}}\equiv~n_D/n_H$             &		 2.68$\times10^{-5}$\\
~~~~~~$X_{^3\mbox{He}}\equiv~n_{^3He}/n_H$          &   1.06$\times10^{-5}$\\
~~~~~~$X_{^4\mbox{He}}\equiv~n_{^4He}/n_H$         &   7.90$\times10^{-2}$\\
~~~~~~$X_{^7\mbox{Li}}\equiv~n_{^7Li}/n_H$          &  ~4.30$\times10^{-10}$\\
\hline
\multicolumn{2}{l}{}\\
\multicolumn{2}{l}{$^a$ $n$ -- is the element number density in the medium.}\\
\multicolumn{2}{l}{\small{$^b$ The data were taken from Steigman (2007). The abundances}}\\
\multicolumn{2}{l}{\small{~~were reconciled with the baryon density $\Omega_B$ obtained}}\\
\multicolumn{2}{l}{\small{~~ by analyzing the CMB anisotropy data.}}
\end{tabular}
\end{table}

Boyd et al. (1989) estimated this effect using the
galactic center of NGC 4151 as an example. The authors
showed that the gamma-ray source associated
with the galactic center could cause $^4$He photodisintegration
and significant D production. However, the
radius of influence of this process is very small and
does not exceed 10~light-years in the case under consideration.
The authors conclude that, given the rarity
of gamma-ray sources in the Galaxy, the change in
the isotopic composition of the Galactic medium is,
in general, unlikely. Nevertheless, Casse et al. (1999)
pointed out that blazars could affect the composition
of the matter accreting onto a black hole. When outflowing
from the blazar’s central region, this matter
can condense into intergalactic clouds. The authors
also note that the final D abundance is sensitive to
the spectral index of the quasar emission. However,
the authors made no quantitative estimations.

The goal of this paper is to consider the mechanism
of element photodisintegration in a medium
by the hard X-ray and gamma-ray emission from quasars by taking into account
various spectral indices of emission and the
high gamma-ray luminosities (compared to the values
used by Boyd et al. 1989).

\section{THE CROSS SECTIONS FOR
PHOTONUCLEAR REACTIONS}
\label{cs}
\noindent

One of the main points of our work is the determination
of photonuclear reaction rates. To calculate the
reaction rates in a medium irradiated by a gamma-ray
flux with various spectral indices $\Gamma$, the dependence of
the reaction cross sections on the gamma-ray energy
should be known.

\begin{figure}
%fig7.
\centering
\includegraphics[width=100mm,clip]{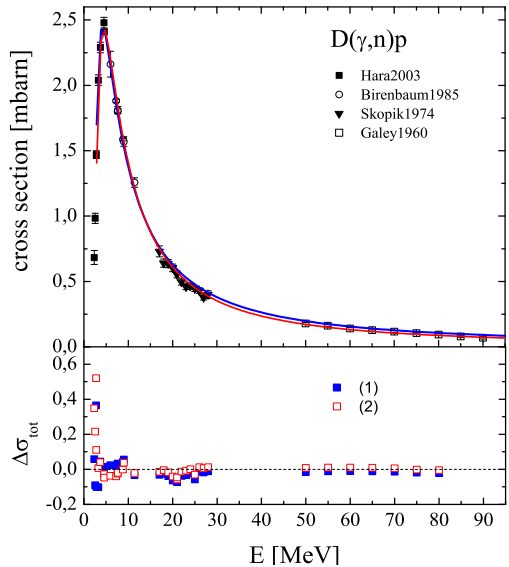} %157
\caption{~Cross section for the D($\gamma$, n)p reaction versus $\gamma$-ray energy E. The dots indicate the experimental data from Hara
et al. (2003), Birenbaum et al. (1985), Skopik et al. (1974), and Galey (1960). The solid curve was computed using Eq. (\ref{eq1}) with
$\sigma_0=19.52$mbarn, $\chi^2=286.0$. The dashed line represents fit (\ref{eq2}) with the following parameters: $\sigma_0=14.3$~mbarn $\beta=1.19$, $\gamma=2.55$, $\chi^2=173.5$.}
\label{CS_D}
\end{figure}

For deuterium (the simplest nuclear system), the
dependence of the interaction cross section on the
gamma-ray energy E can be calculated analytically
(Bethe and Peierls 1935) and is written as:
\begin {equation}
\label{eq1}
\sigma(E)=\sigma_0Q^{3/2}\frac{\left({E-Q}\right)^{3/2}}{E^3}
\end {equation}
where $\sigma_0$  is a dimensional constant, and $Q$ is the reaction
threshold. This dependence of the cross section
on the gamma-ray energy describes well the experimental
data on deuterium photodisintegration (see
Fig.~\ref{CS_D}).

\begin{figure*}
%fig7.
\centering
\includegraphics[width=157mm,clip]{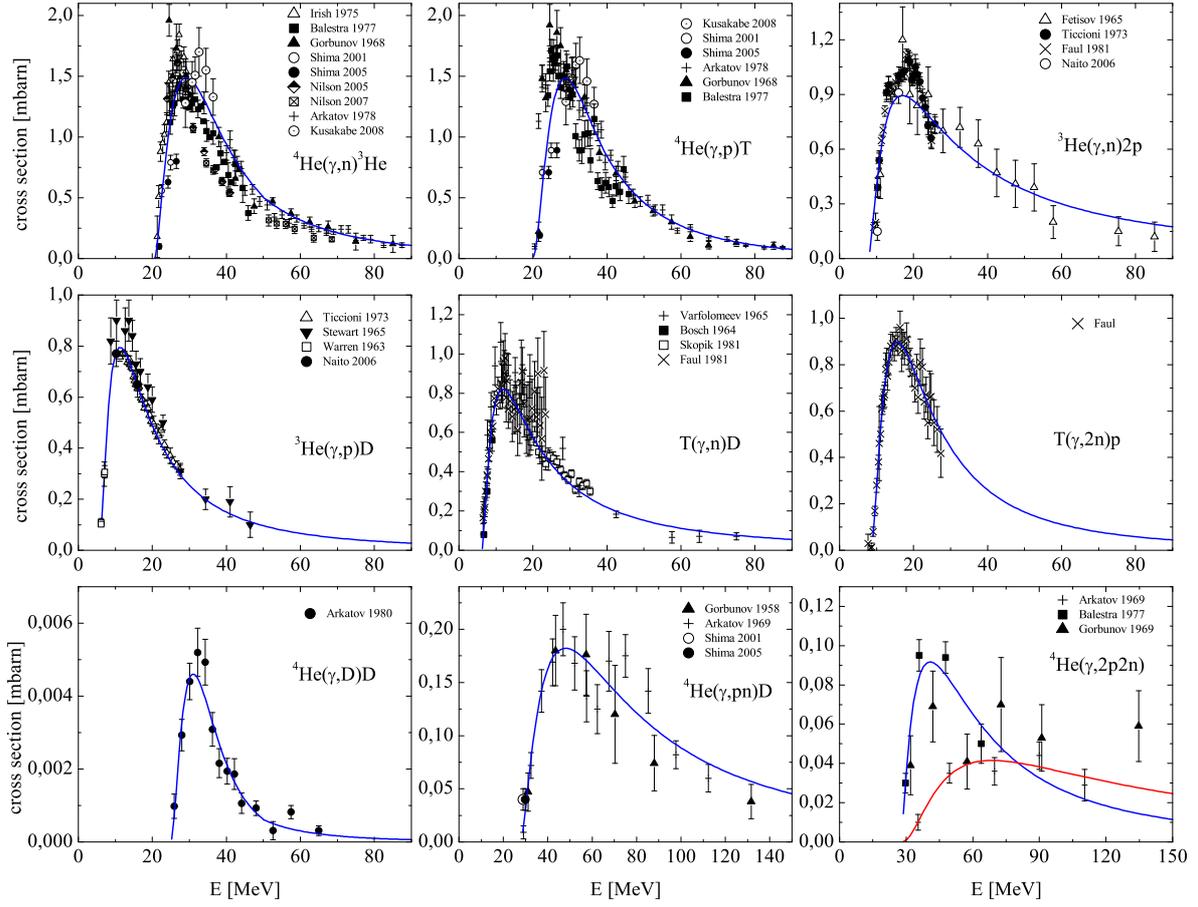} %157
 \caption{~~Photonuclear reaction cross sections versus $\gamma$-ray energy E. The solid curves represent fit (\ref{eq2}) with  $\sigma_0, \beta, \gamma$ from Table~\ref{table2}. The dots indicate the experimental data:
T$\left(\gamma,n\right)$D~Varfolomeev~and~Gorbunov~(1965), Bosch~et~al.~(1964), Skopik~et~al.~(1981), Faul~et~al.~(1981); T$\left(\gamma,2n\right)$p~Faul~et~al.~(1980).
$^3$He$\left(\gamma,p\right)$D~Ticcioni~et~al.~(1973), Stewart~et~al.~(1965), Warren~et~al.~(1963), Naito~et~al.~(2006);
$^3$He$\left(\gamma,n\right)$2p~Fetisov~et~al.~(1965), Ticcioni~et~al.~(1973), Faul~et~al.~(1981), Naito~et~al.~(2006);
$^4$He$\left(\gamma,p\right)$T~Shima~et~al.~(2001,2005), Kusakabe~et~al.~(2009), Arkatov~et~al.~(1978b), Gorbunov~(1968), Balestra~et~al.~(1977);
$^4$He$\left(\gamma,n\right)^3$He~Irish~et~al.~(1975), Balestra~et~al.~(1977), Gorbunov~(1968), Shima~et~al.~(2001,2005), Kusakabe~et~al.~(2009), Nilson~(2005,2007), Arkatov~(1978a);
$^4$He$\left(\gamma,D\right)$D~Arkatov~et~al.~(1980);
$^4$He$\left(\gamma,pn\right)$D~Shima~et~al.~(2001,2005), Arkatov~et~al.~(1969b), Gorbunov~(1958);
$^4$He$\left(\gamma,2p2n\right)$~Balestra~et~al.~(1977), Arkatov~et~al.~(1969a), Gorbunov~(1969).}
 \label{cs_other}
\end{figure*}

Other elements have a more complex nuclear
structure, and the theoretical calculations for many
of them are not in satisfactory agreement with the
experimental data (Dzhibuti et al. 1965; Balestra
et al. 1977). Figure~\ref{cs_other} presents the experimental
photonuclear reaction cross sections and their fits.
The dependence of the reaction cross sections for
other elements on the gamma-ray energy is similar
to the deuterium reaction cross section: there exist
a threshold energy starting from which the reaction
proceeds and a power-law decrease in the cross section.
Taking this into account, we chose functions of
the following form for the fit:
\begin {equation}
\label{eq2}
\sigma(E)=\sigma_0\frac{\left({\epsilon-1}\right)^\beta}{\epsilon^\gamma},
\end{equation}
where $\sigma_0$, $\beta$ and $\gamma$ are the parameters of the fit, and
 $\epsilon\equiv E/Q$ is the dimensionless energy. The best-fit
parameters are given in Table 2. For three $^4$He destruction
reactions, the cross section near the threshold
and at high energies cannot be described by one
function. Therefore, for a more accurate fit, the energy
range was divided into two parts, each of which was
fitted independently. For the $^4He(\gamma,2p2n)$ reaction,
there is a discrepancy in the absolute value of the
cross section and the position of the maximum between
the data from Arkatov et al. (1969, 1970), Gorbunov (1969),
 and Balestra et al. (1977). As was
pointed out by Arkatov et al. (1970), this is because
some arbitrariness is admitted when separating the
individual events of the $(\gamma,pn)$ and $(\gamma,2p2n)$ reactions.
In this case, the cross section was described by
two functions, one of which was constructed from the
data by Arkatov~et~al.~(1970) and the other function
was constructed from the data by Gorbunov (1969)
and Balestra et al. (1977). However, the contribution
from the $^4He(\gamma,2p2n)$ reaction to the change in the
abundance of the elements is so small that the difference
between the two fits does not lead to a noticeable
change in the final result.

\begin{table*}
    \caption{Parameters of the fit to the photonuclear cross sections\,$^{a}$.}
    \label{table2}
    \centering
%\vspace{6mm} \centering {{\bf Table 2.} Parameters of the fit to the photonuclear cross sections\,$^{a}$.}\label{table2}
\begin{tabular}{l|c|c|c|c|c}
 \hline\hline
 Reaction~~~~  &      Q, MeV      &       ~~E$\gamma$, MeV      &   $\sigma_0$, mbarn      & ~~~~~$\beta$~~~~~ & $\gamma$ \\
\hline
${\rm \;\,D}(\gamma,n)p            $   & 2.23  & & 14.3 & 1.19 & 2.55 \\[5pt]

${\rm \;\,T}(\gamma,n){\rm D}      $   & 6.257 & & 9.86 & 1.7   & 3.6   \\
${\rm \;\,T}(\gamma,2n)p           $   & 8.482 & & 22.1 & 2.11  & 4.65  \\[5pt]

$^3{\rm He}(\gamma,p){\rm D}       $   & 5.493 & & 21.4 & 2.43  & 4.75  \\
$^3{\rm He}(\gamma,n)2p            $   & 7.72  & & 9.48 & 1.87  & 3.43  \\[5pt]

$^4{\rm He}(\gamma,p){\rm T}       $   & 19.81 & $\left\{
\begin{tabular}{c}$E_{\gamma}<40$\\$E_{\gamma}>40$\\ \end{tabular}
\right. $ &\begin{tabular}{c} 145.5\\ 19.5\\ \end{tabular} & \begin{tabular}{c} 2.3\\ 1.0 \\ \end{tabular}
& \begin{tabular}{c} 7.4\\ 4.5 \\ \end{tabular} \\

$^4{\rm He}(\gamma,n)^3{\rm He}    $   & 20.57 & $\left\{
\begin{tabular}{c}$E_{\gamma}<50$\\$E_{\gamma}>50$\\ \end{tabular}
\right. $ &\begin{tabular}{c} 60.2\\ 9.11\\ \end{tabular} &
\begin{tabular}{c} 1.8\\ 0.6 \\ \end{tabular}
& \begin{tabular}{c} 6.1\\ 3.5 \\ \end{tabular} \\

$^4{\rm He}(\gamma,{\rm D}){\rm D} $   & 25.0 & $\left\{
\begin{tabular}{c}$E_{\gamma}<50$\\$E_{\gamma}>50$\\ \end{tabular}
\right. $ &\begin{tabular}{c} 0.64\\ 0.01\\ \end{tabular} &
\begin{tabular}{c} 1.95\\ 0.1 \\ \end{tabular}
& \begin{tabular}{c} 10.0\\ 4.3 \\ \end{tabular} \\

$^4{\rm He}(\gamma,pn){\rm D}      $   & 28.0 & & 2.1 & 1.5 & 3.6 \\

$^4{\rm He}(\gamma,2p2n)           $   & 28.0 & & 0.81 & 1.1 & 3.5 \\

\hline
%\multicolumn{6}{l}\\ %[-4mm]
\multicolumn{6}{l}{$^a$ The fitting formulas give an accuracy of at least 10\%.}\\
\multicolumn{6}{l}{Note. The spontaneous decay rate of tritium ${\rm T}(\beta^-)^3{\rm He}$ is $k=1.82709\pm0.0026\times10^{-9}\,~s^{-1}$, see Akulov and Mamyrin (2004).}\\
\end{tabular}
\end{table*}

Given the high gamma-ray energy, the photodisintegration
reaction products can have a kinetic energy
exceeding their binding energy. In this case, the
thermalization of high-energy D and $^3$He nuclei in the
medium should be considered separately. Mostly the
nuclei will collide with cold protons and will reduce its
energy or be destroyed, depending on the energy of the
colliding particles and the scattering pattern (elastic,
inelastic). Deuterium has a low binding energy,
Q$_D$=2.23~MeV, will be most sensitive to the thermalization
process. The width of the reaction cross section
maximum is  $\Delta_E\sim2\div10$~MeV (see Fig.~\ref{cs_other} and
Table~\ref{table2}). The kinetic energy of the produced nuclei
can be estimated as $E_{k}\simeq{\Delta_E/2}$, which can exceed
the deuterium binding energy Q$_D$. Therefore, the
thermalization of D and $^3$He nuclei in themedium can
partially destroy the produced amount of elements.

Thus, complex statistical calculations of the thermalization
of nuclei required to reliably estimate the
amount of produced D and $^3$He nuclei. In this paper,
we consider the effect that consists in a great change
in the abundance of light nuclei in the medium irradiated
by emission with various values of the spectral
hardness $\Gamma$. This effect turns out to be especially
important in the immediate vicinity of AGNs.

\section{OBSERVATIONAL DATA ON ACTIVE
GALACTIC NUCLEI}
\label{Data}
\noindent
Blazars and quasars (AGNs) are extremely intense
sources of energy release in the Universe; they
also have enormous gamma-ray luminosities. Since
gamma rays with an energy above 2.23~MeV (the
deuterium binding energy) are involved in the photodisintegration
of elements, we are interested in the
gamma-ray energy range from 1 to 1000 MeV. The
upper boundary was chosen with some margin: since
the quasar spectrum falls off as a power law, taking
into account the energy range from 100 to 1000 MeV
causes the reaction rates to increase by no more
than 1\%.

The spectra of such objects were investigated at
the orbital observatories: AGILE (30-–500 MeV; see,
e.g., D'Ammando et al. 2011), COMPTEL (0.75–-30 MeV; see, e.g., Schonfelder et al. 2000), EGRET
(30 MeV–-20 GeV; see, e.g., Hartman et al. 1999),
and Fermi LAT (20 MeV–-300 GeV; see, e.g., Abdo
et al. 2010).

\begin{figure*}
%fig7.
\centering
\includegraphics[width=100mm,clip]{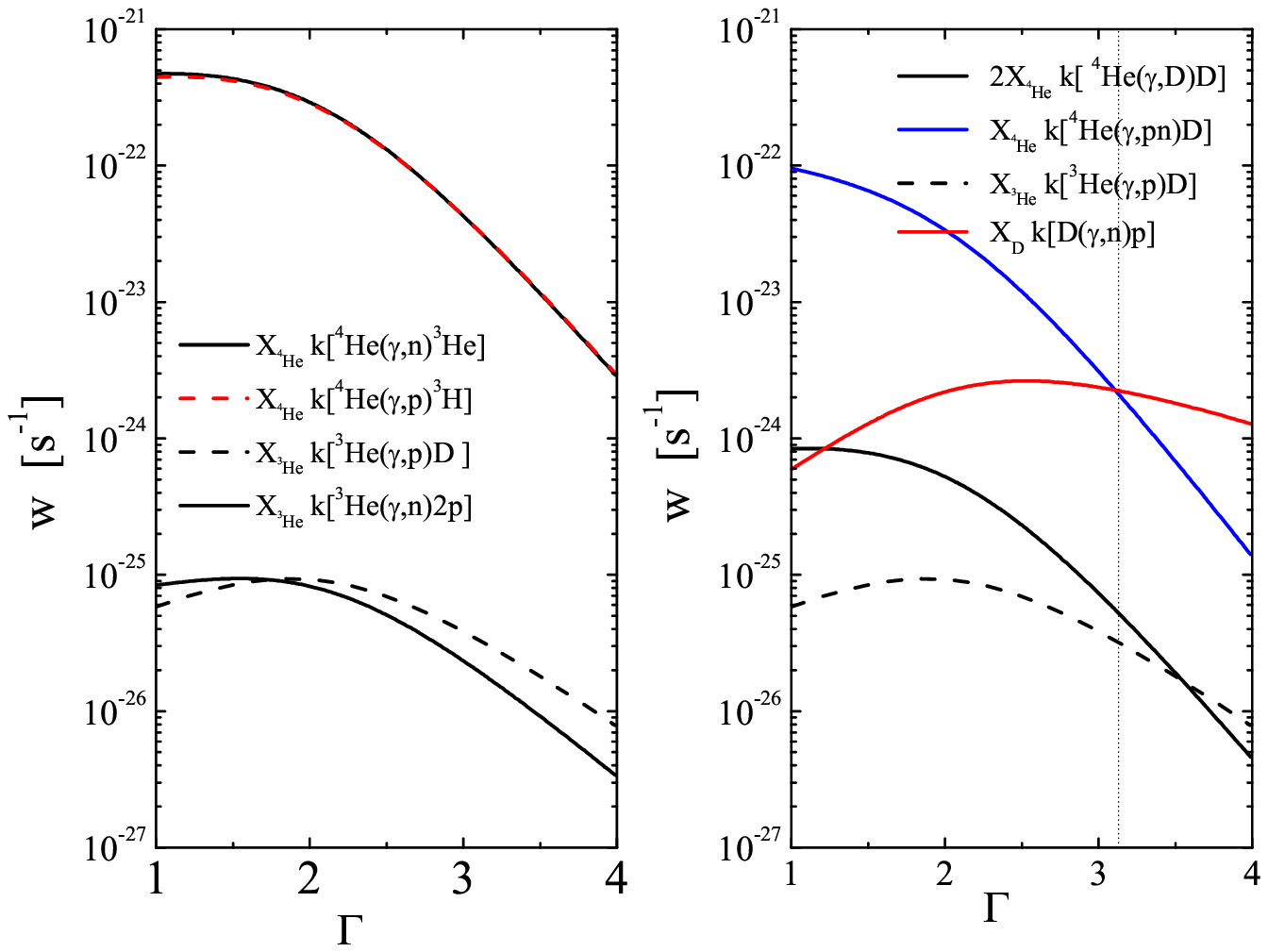} %157
 \caption{~~Effective D, $^3$He and $^4$He production and destruction rates when primordial matter is exposed to $\gamma$-rays with energy E = 1--1000 MeV
with the spectrum $dN/dE=N_0(E_0/E)^{\Gamma}$ versus  $\Gamma$. $N_0$ corresponds to the $\gamma$-ray flux with L$_{\gamma}=10^{47}$~erg~s$^{-1}$
at the distance of 1 kpc.}
 \label{fig03}
\end{figure*}

In the region of photonuclear reaction thresholds
(2-–30~MeV), there are data on the spectra from the
survey by Schonfelder et al. (2000). As was shown in
the surveys by Hartman et al. (1999) and Schonfelder~et~al.~(2000),
 the quasar emission at high energies
($E_\gamma~>~1~$MeV) is well fitted by a power law:
\begin{equation}
\label{eq3}
F_{\gamma}(E)=F_0\left(\frac{E}{\varepsilon_0}\right)^{-\Gamma}
\end{equation}
where F$_0$ is a flux density in phot~cm$^{2}$~s$^{-1}$~MeV$^{-1}$,
$\Gamma$ is the spectral index, and $\varepsilon_0$ -- is the scale factor.
Their typical values are $F_0\sim10^{-8}$~phot~cm$^{2}$~s$^{-1}$~MeV$^{-1}$,
$\Gamma=1\div4$, $\varepsilon_0=1\div300$~MeV.

\section{KINETICS OF THE ISOTOPIC
COMPOSITION OF THE INTERSTELLAR
MEDIUM}
\label{Kinetic}
\begin{figure*}
%fig7.
\centering
\includegraphics[width=130mm,clip]{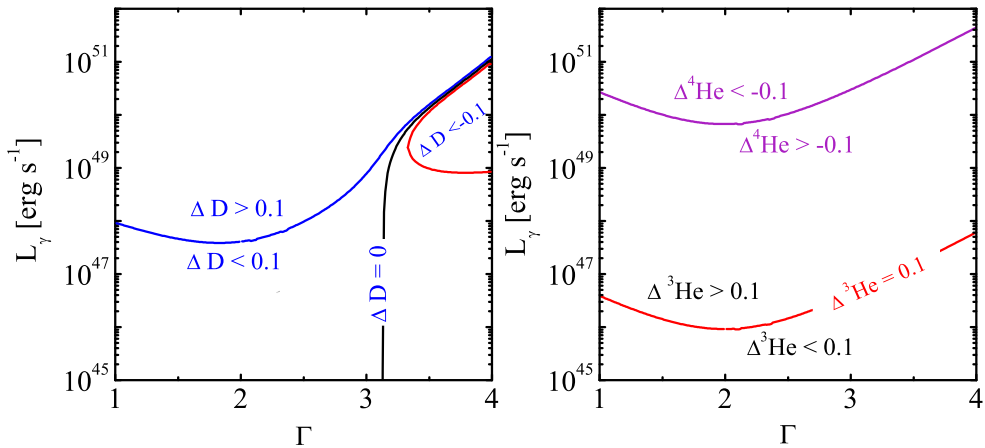} %100
 \caption{~~Total luminosity of the $\gamma$-ray source L$_{\gamma}$ in the energy range 1–-10$^3$~MeV versus spectral slope $\Gamma$ corresponding to a
$\pm 10$\% change in the abundance of D and $^3$He in the primordial matter at a distance of 1 kpc in an exposure time of 10$^9$~yr.}
 \label{l_gamma}
\end{figure*}

\begin{figure*}
%fig7.
\centering
\includegraphics[width=100mm,clip]{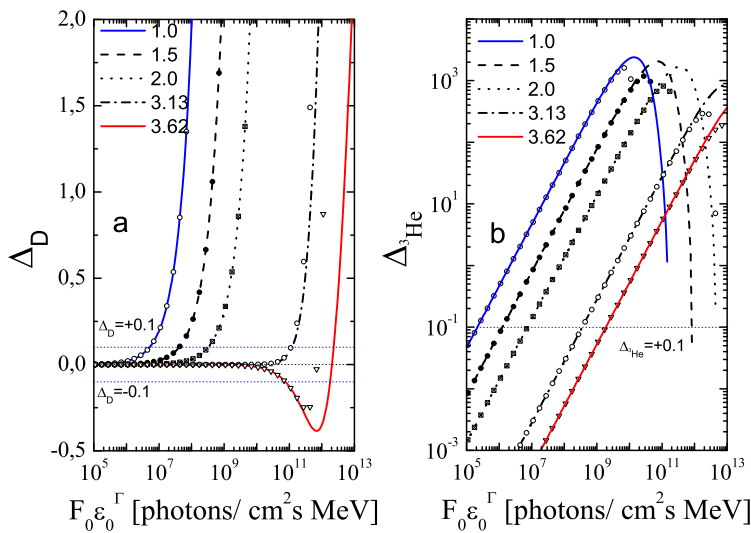} %157
 \caption{~~Change of the D (a) and $^3$He (b) abundances in a medium exposed for $10^9$ yr versus flux density at an energy of 1~MeV
$F_0\varepsilon_0^{\Gamma}$. As the spectral slope $\Gamma$ increases, the curve is shifted rightward; at $\Gamma > \Gamma_{\mbox{k}}=3.13$, the $\Delta_D$ curve acquires a negative
value (D is destroyed). $\Gamma=3.62$ corresponds to the minimum flux density $F_0\varepsilon_0^{\Gamma}\simeq8\times10^{11}$~phot~cm$^{-2}$~s$^{-1}$~MeV$^{-1}$ at which
10\% of D will be destroyed in 10$^9$ yr.}
 \label{delta_flux}
\end{figure*}

\begin{figure*}
%fig7.
\centering
\includegraphics[width=100mm,clip]{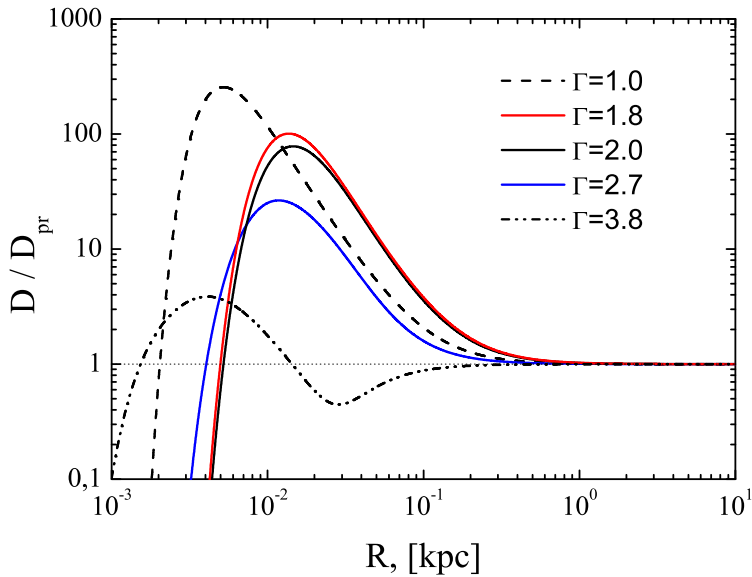} %157
 \caption{~~
D/H ratios in the medium versus distance to a quasar of luminosity L$_{\gamma}=10^{47}$~erg~s$^{-1}$. The D abundance relative to its primordial value (D/H)$_{pr}$$=2.68\times10^{-5}$ is along the vertical axis. The curves are presented for various spectral slopes $\Gamma$ = 1.0, 1.5, 1.8, 2.7, 3.8 of a source with the same luminosity. $\Gamma$ = 2.7 corresponds to the spectral slope for the quasar
emission considered by Boyd et al. (1989). The largest amount of D is produced at $\Gamma$ = 1.83.}
 \label{deut_radius}
\end{figure*}

\noindent
The kinetics of the isotopic composition of the
interstellar medium as a function of the exposure time
and spectral parameters can be obtained by solving
the system of ordinary differential equations (\ref{eq4} - \ref{eq7}):
\begin{equation}
    \begin{split}
        \label{eq4}
        &\frac{dX_{^4He}}{dt}=-(k_{^4He\to T}+k_{^4He\to ^3He}+{}\\
        &+k_{^4He\to D+D}+k_{^4He\to D+pn}+k_{^4He\to2p+2n})X_{^4He}
    \end{split}
\end{equation}
\begin{equation}
\begin{split}
\label{eq5}
&\frac{dX_{T}}{dt}=-\left({k_{T\to ^3He}+k_{T\to D}+k_{T\to
2n+p}}\right)X_{T}+{}\\
&+{k_{^4He\to{^3H}}}{X_{^4He}}
\end{split}
\end{equation}
\begin{equation}
\begin{split}
\label{eq6}
&\frac{dX_{^3He}}{dt}=-\left({k_{^3He\to
D}+k_{^3He\to
n+2p}}\right)X_{^3He}+{}\\
&+{k_{^4He\to{^3He}}}{X_{^4He}}+{k_{T\to{^3He}}}{X_{T}} %{}\\
\end{split}
\end{equation}
\begin{equation}
\begin{split}
\label{eq7}
&\frac{dX_D}{dt}=-{k_{D\to p+n}}X_D+(2\cdot
k_{^4He\to D+D}+{}\\
&+k_{^4He\to D+pn}){X_{^4He}}+k_{^3He\to D}X_{^3He}+{k_{T\to D}}{X_{T}},
\end{split}
\end{equation}
where $X$ denote the elemental abundances, and $k$
are the photonuclear reaction rates. The element
abundances calculated using the theory of Big Bang
nucleosynthesis with the baryon density $\Omega_B$ obtained
by analyzing the cosmic microwave background
(CMB) anisotropy data (see Table~\ref{table1}) were taken as
the medium's initial composition. The influence of
the elements with an atomic weight greater than that
of $^4$He on the change in the abundances of lighter
elements may be neglected, because their relative
abundance in a medium similar to the primordial one
is low (($^7$Li/H)$\le5\times10^{-10}$), while the cross sections
are of the same order of magnitude (see Berman
et al. 1965; Varlamov et al. 1986). Using Eqs.~(\ref{eq2})
and~(\ref{eq3}), the photonuclear reaction rates can be
written as:
\begin{equation}
\label{eq8}
\begin{split}
&k=\int\limits_Q^\infty \sigma \left ( E \right)  F_\gamma(E) \, dE=10^{-27}\left(\frac{\sigma_0}{1~\mbox{mbarn}}\right)\times{}\\
&\times F_0{\varepsilon_0}
^{\Gamma}\left(\frac{Q}{1~\mbox{MeV}}\right)^{1-\Gamma}\int\limits_1^{1000/Q}\frac{{\left(x-1\right)}^\beta}{x^{\gamma+\Gamma}} \,dx,
\end{split}
\end{equation}
where the dependence on $\beta,~\gamma$ and $\Gamma$ is determined by
an integral that can be expressed in terms of Euler’s incomplete beta functions:
\begin{equation}
\begin{split}
\label{eq9}
&\int\limits_1^{1000/Q}\frac{{\left(x-1\right)}^\beta}{x^{\gamma+\Gamma}} \,dx=
\mbox{B}_1(\Gamma+\gamma-\beta-1,\beta+1)+{}\\
&+\mbox{B}_{Q/1000}(\Gamma+\gamma-\beta-1,\beta+1).
\end{split}
\end{equation}
The system of equations (\ref{eq4})--(\ref{eq7}) is solved analytically.
In view of the fitting formulas (\ref{eq8}) and (\ref{eq9}), the element
abundances at any instant of time are expressed in
algebraic form in terms of special functions. This
allows the distributions of elemental abundances in
the medium to be rapidly calculated for various spectral
parameters. The results of our calculations are
presented in {Figures~{\ref{l_gamma}--\ref{balbes}}}.

The change of the element abundances for a fixed
spectral slope $\Gamma$ is determined by only one parameter
$x=F_0\varepsilon_0^{\Gamma} t$, the multiplication of the gamma-ray flux
at an energy of 1~MeV by the exposure time. We assume
that the gamma-ray flux is produced by the central
source and is determined only by the distance to
the quasar and its luminosity. Taking into account the
correlation between the optical and gamma-ray luminosities
(Arshakian 2011), we may consider the optical
Eddington luminosity L$_{Edd}=4\pi{GM}m_{p}c/\sigma_T=1.5\times10^{46}$~M$_8$~erg~s$^{-1}$,
 where $\sigma_T$ is the Thomson cross section,
 as an upper limit for the gamma-ray
luminosity of sources. The range of admissible cloud
exposure times is estimated via the lifetime of a quasar
at its active phase t$ = 10^5\div 10^9$~yr (Martini and Weinberg
2001). We consider the characteristic exposure
time t$_0 = 10^9$~yr to be able to compare our results with
the work by Boyd et al. (1989) and to take into account
the maximally possible effects when estimating
the deuterium production.

For the limiting gamma-ray luminosity {L$_{\gamma}=10^{47}$~erg~s$^{-1}$}
(i.e., the Eddington luminosity of a quasar with a mass of $10^9~M_{sun}$),
the element abundances changes by $\simeq10\%$ at distances of $\sim100$ pc.
The self-similarity of the model allows the radius
of influence of the quasar (on the medium's composition)
to be recalculated for such quasars of a
different luminosity and for a different exposure time
by rescaling:
\begin{equation}
\label{eq10}
R_{1} = R_{2}\left(\frac{L_{1}t_{1}}{L_{2}t_{2}}\right)^{(1/2)}.
\end{equation}

In contrast to previous works, we also investigated
the dependence of the reaction rates on the spectral
index. For a quasar with a fixed luminosity  L$_{\gamma}=10^{47}$~erg~s$^{-1}$
in the energy range 1--1000 MeV, the D and $^3$He production
 and destruction rates are plotted
against the spectral hardness $\Gamma$ in Fig.~\ref{fig03}. These plots
describe the dynamics of the medium's composition
on short exposure time scales, when the relationship
between the abundances of $^4$He and $^3$He, D in the
medium did not change significantly. The maximum
in the reaction rates (at $\Gamma\simeq 2$) is determined by the
relationship between the position of the maximum in
the reaction cross sections and the power-law shape
of the spectrum. For deuterium (Fig.~\ref{fig03}b), there is a
feature $\Gamma=3.1$. For harder spectra ($\Gamma\le3.1$), deuterium
in the medium will be only produced (through
the photodisintegration of $^4$He); conversely, for softer
spectra ($\Gamma\ge3.1$), the destruction of deuterium can
exceed its production. There are no such features for
$^3$He (see Fig.~\ref{fig03}a).it is only produced. The contributions
from the $^3$He production during the $^4$He
destruction exceeds the $^3$He photodisintegration by
more than two orders of magnitude. Since there are
no $^4$He sources in the model under consideration, $^4$He
is only destroyed. The amount of tritium is established
at a constant level (until the $^4$He abundance changes
significantly) determined by the equilibrium condition
$dX_{T}/dt=0$:
\begin{equation}
\label{eq11}
X_{T}=X_{^4He}\frac{k_{^4He\to T}}{K_{T}}=X_{^4He}(0)\frac{k_{^4He\to T}}{K_{T}}\exp(-K_{^4He}t),
\end{equation}
where $K_{^4He}$ and $K_{T}$ denote the total element destruction
rates.

The features in the relationships between the reaction
rates determine the pattern of changes in the
element abundances. Figure~\ref{l_gamma} define the regions of
spectral parameters (the gamma-ray luminosity of
the quasar $L_{\gamma}$ for an isotropic source and the spectral
index $\Gamma$) corresponding to a 10\% change of the
elemental abundances in a cloud $\Delta_{\mbox{D}}$ and $\Delta_{^3\mbox{He}}$ at
a distance of 1~kpc in an exposure time of 1 Gyr.
As has been pointed out above, the destruction of
D is possible in a certain domain of parameters. In
Fig.~\ref{delta_flux}, the change in the abundances of D and $^3$He
is plotted against the spectral flux density $F_0\varepsilon_0^{\Gamma}$
at an energy of 1~MeV for various spectral indices $\Gamma$. The
abundance of D and $^3$He is determined only by the
exposure parameter  $x\propto F_0\cdot t$. Therefore, apart from
the spatial change in elemental abundances (the flux
density decreases as $F_0\propto r^{-2}$), these plots illustrate
the temporal evolution of the D and $^3$He abundances
in the medium.

Another feature in the distribution of deuterium is
that when the medium is irradiated by emission with
a soft spectrum, the decrease in the mass fraction
of D is subsequently (at high flux densities or on
long exposure time scales) replaced by a significant
increase. This is because a new D production channel
appears from the newly produced $^3$He.

\section{THE EFFECTIVE RADIUS OF INFLUENCE. AN ISOTROPIC SOURCE}
\label{Efrad}
\noindent
The effective radius of influence of a quasar as a
function of its gamma-ray luminosity and spectral
hardness in a time of $\sim$1~Gyr can be determined within
the spherically symmetric quasar emission model.
As has been shown above, the characteristic radius
of influence of a quasar with a luminosity L$_{\gamma}=10^{47}$~erg~s$^{-1}$
in the energy range 1--1000 MeV is $\sim$ 100 pc. The Hubble expansion may be neglected
at such distances, and the flux density is then related
to the luminosity by the standard relation
\begin{equation}
\label{eq12}
\frac{L_\gamma}{4\pi R^2} = \int\limits_1^{1000} F_\gamma(E) E \, dE =  F_0\varepsilon_0^\Gamma I(\Gamma),
\end{equation}
the latter expression was derived by taking into account
Eq. (\ref{eq3}), where $I(\Gamma)=\frac{1000^{2-\Gamma}-1}{2-\Gamma}$ and
$I(\Gamma)=\ln(1000)$ at $\Gamma=2.0$.

\begin{figure*}
%fig7.
\centering
\includegraphics[width=100mm,clip]{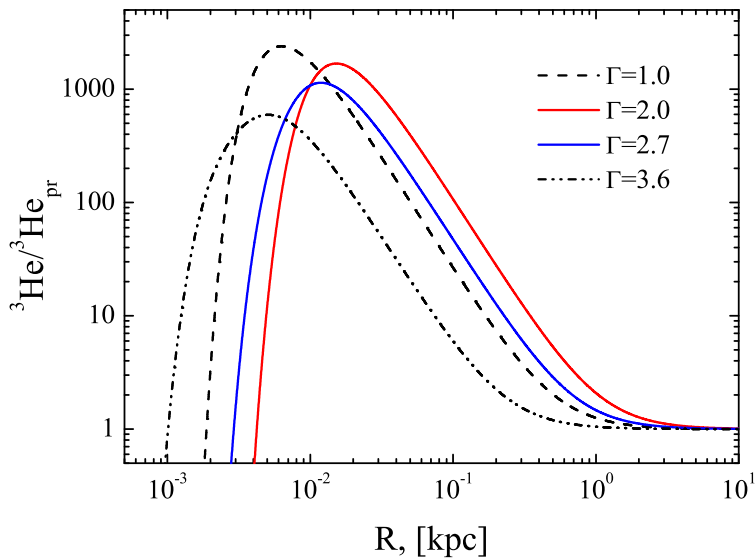} %157
 \caption{~~~$^3$He/H ratio in the medium versus distance to a quasar of luminosity L$_{\gamma}=10^{47}$~erg~s$^{-1}$. The $^3$He abundance relative
to its primordial value ($^3$He/H)$_{pr}=1.06\times10^{-5}$ is along the vertical axis. The curves are presented for various spectral slopes
$\Gamma$= 1.0, 2.0, 2.7, 3.6 of a source with the same luminosity. $\Gamma$ = 2.7 corresponds to the spectral slope of the quasar emission
considered by Boyd et al. (1989). $\Gamma$ = 2.0, the maximum amount of $^3$He is produced in the medium.}
 \label{he3_radius}
\end{figure*}

Figures \ref{deut_radius} and \ref{he3_radius} show the profiles of the D/H
and $^3$He/H ratios as a function of the quasar distance.
The distribution of elements in the medium
changes greatly as the quasar is approached. The
main change occurs at $\sim$100 pc, where the amount
of D and $^3$He increases by two or three orders of
magnitude. At shorter distances, effective photodisintegration
of both $^4$He and D and $^3$He takes place.
Since there are no $^4$He sources in the model under
consideration, the $^4$He abundance decreases exponentially
with increasing flux density or decreasing
quasar distance. Thus, the effect of a change in the D
and $^3$He abundances is strongly local (within $\sim$1 kpc
of the AGN center).

The changes in the element abundances are also
sensitive to the spectral hardness $\Gamma$. For different
spectra, the elemental abundances in the medium can
differ by an order of magnitude at the same exposure
time and quasar luminosity. Figures~\ref{deut_radius} and~\ref{he3_radius} present
several curves corresponding to different spectral
slopes. The dependence of the radius of influence on
the spectral index $\Gamma$ is determined by the dependence
of the reaction rates (see Fig.~\ref{fig03}): D and $^3$He are
produced in large quantities and at great distances
at $\Gamma \simeq 1.8$ and 2.0, respectively.

\section{MOTION OF THE MEDIUM}
\label{motion}
\noindent

The distributions of the abundances of light elements
formed through photonuclear reactions (see
Figs. \ref{deut_radius} and \ref{he3_radius}) were obtained by assuming the
medium to be at rest, when the matter at different distances
from the quasar center is not mixed. In reality, the
matter in the interstellar medium moves, and the
distributions of the elements change.

Estimating the global effect of the contribution
from the emission of quasars to the change of the
D and $^3$He abundances in the Universe as a whole
(after the completion of Big Bang nucleosynthesis),
we perform a uniform averaging of the changed D and $^3$He
 abundances over the space in which the change
occurred:
\begin{equation}
\label{eq13}
<\mbox{X}>(R,L_{\gamma},\Gamma)=\frac{\overline{n}}{n_{pr}}=\frac{1}{V}\frac{\int_0^R X(r)4\pi r^2 dr}{X_{pr}},
\end{equation}
where X denotes the abundances of the elements:
(D/H), ($^3$He/H) and ($^4$He/H).

\begin{figure*}
%fig7.
\centering
\includegraphics[width=100mm,clip]{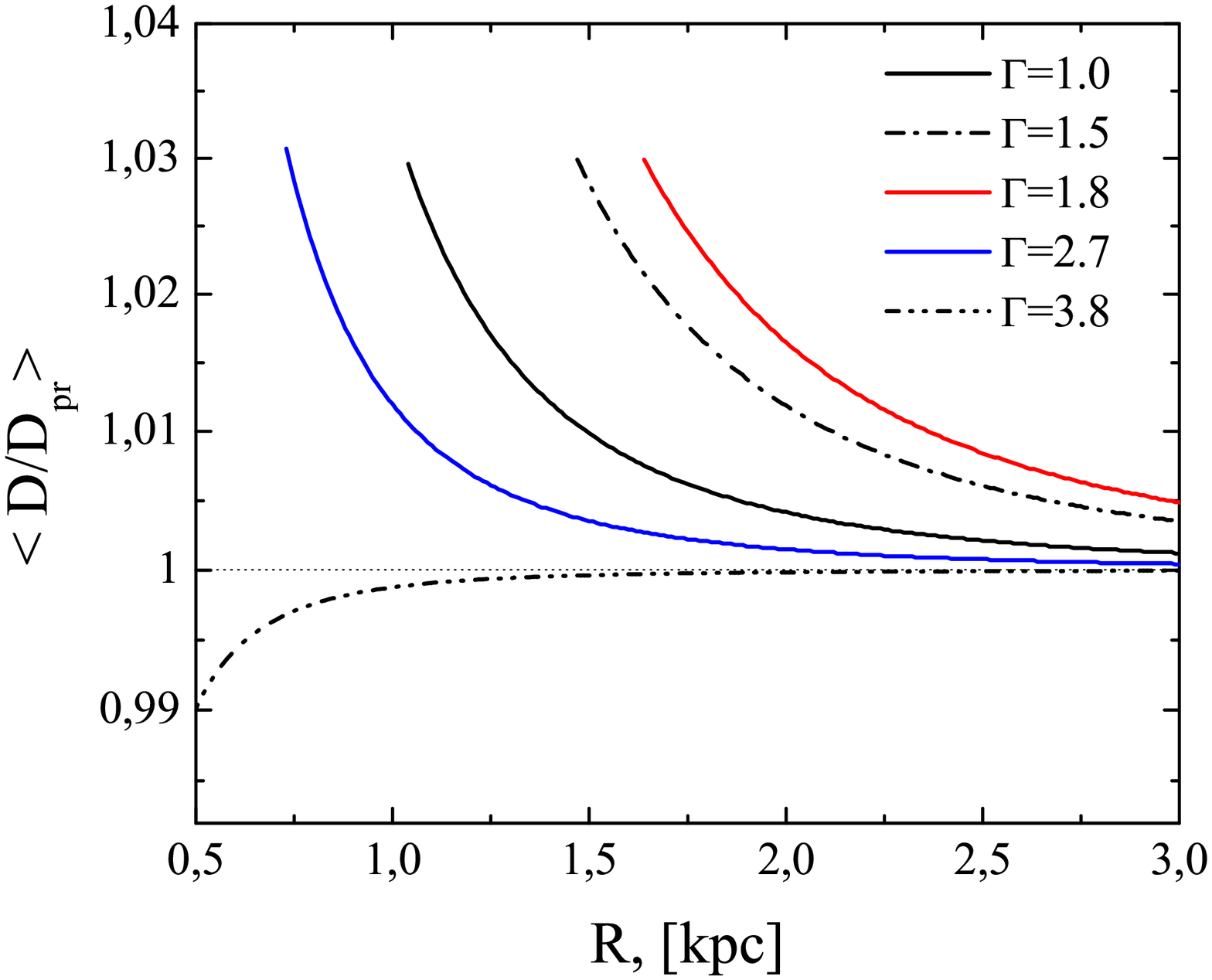} %157
 \caption{~~Average D/H ratio versus size of the averaging region corresponding to a change in the composition of primordial
matter when exposed to $\gamma$-rays with an energy of 1–1000 MeV from a source of luminosity L$_\gamma=10^{47}$~erg~s$^{-1}$. D/H in units of
its primordial value (D/H)$_{pr}$ is along the vertical axis. The curves are presented for several spectral slopes $\Gamma$=1.0, 1.5, 1.83, 2.7, 3.8.
$\Gamma$ = 1.83 corresponds to the maximum amount of D produced in the medium. Boyd et al. (1989) used the model of a
source with a spectral slope $\Gamma$ = 2.7.}
 \label{deut_average}
\end{figure*}

\begin{figure*}
%fig7.
\centering
\includegraphics[width=100mm,clip]{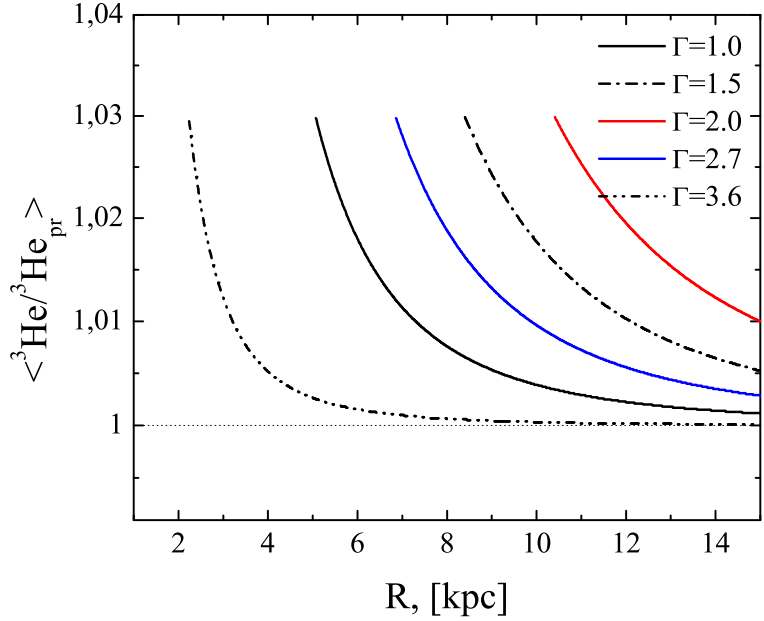} %157
 \caption{~~~~Average ($^3$He/H) ratio versus size of the averaging region corresponding to a change in the composition of primordial
matter when exposed to $\gamma$-rays with an energy of 1–1000 MeV from a source of luminosity L$_\gamma=10^{47}$~erg~s$^{-1}$. $^3$He/H in units
of its primordial value ($^3$He/H)$_{pr}$ is along the vertical axis. The curves are presented for several spectral slopes $\Gamma$ = 1.0, 1.5,
2.0, 2.7, 3.6. $\Gamma$ = 2.0 corresponds to the maximum amount of $^3$He produced in the medium. Boyd et al. (1989) used the model
of a source with a spectral slope $\Gamma$ = 2.7.}
 \label{he3_average}
\end{figure*}

The upper limit of the integral R in Eq.~(\ref{eq13}), which
we call the effective zone of influence of a quasar, is
chosen from the conditions under which the quasar
influence will change the composition of the medium
by 10\% at 1 Gyr. The limiting quasar luminosity
L$_{\gamma}=10^{47}$~erg~s$^{-1}$ and exposure time t=$10^9$~yr are
used to estimate the maximum influence of the quasar
emission on the composition of the medium. The size
of the region of influence is $\sim$1~kpc for D and $\sim 10$~kpc
for $^3$He. The deviation of the averaged abundance of
D and $^3$He from its primordial value does not exceed
5\%; the produced D and $^3$He are plotted against the
quasar distance for various spectral slopes in Figs.~\ref{deut_average}~and~\ref{he3_average}.
 The element abundance averaged over a larger
volume decreases in inverse proportion to the distance
squared, $\mbox{$<X>\propto R^{-2}$}$.

Thus, the global effect of the gamma-ray emission
from quasars (at a limiting isotropic luminosity
of $10^{47}$~erg~s$^{-1}$) on the composition of the medium
turns out to be insignificant, and the emission from
quasars cannot be considered to be a source of the
observed excesses of the D and $^3$He mass fractions
above the primordial values.

However, this estimate obtained by assuming uniform
mixing is only a lower limit for the influence of
a quasar on the change in the isotopic composition
of the surrounding medium. Considering the model
of nonuniform mixing, a situation is possible where a
cloud of gas overrich in D and $^3$He can be accelerated
by a jet and be ejected into the medium surrounding
the quasar. The matter from the interstellar medium
can also fall to the quasar through an accretion disk,
be exposed, and then ejected back. The existence
of such a mechanism was previously pointed out by
Casse et al. (1999).

The pollution of the medium through this mechanism
depends on many parameters, such as the
relationship between the required matter exposure
time near the quasar and the cloud lifetime near the
galactic nucleus, the degree of destruction of the
produced elements during the ejection of matter into
the intergalactic medium, and the condensation of the
ejected matter into small intergalactic clouds. This
mechanism seems possible, but a detailed study of the
theory of accretion disks around AGNs, the theory of
jets, and other theories is needed for its assessment.

\section{DISCUSSION}
\label{DISCUSSION}
\noindent

\begin{figure*}
%fig7.
\centering
\includegraphics[width=100mm,clip]{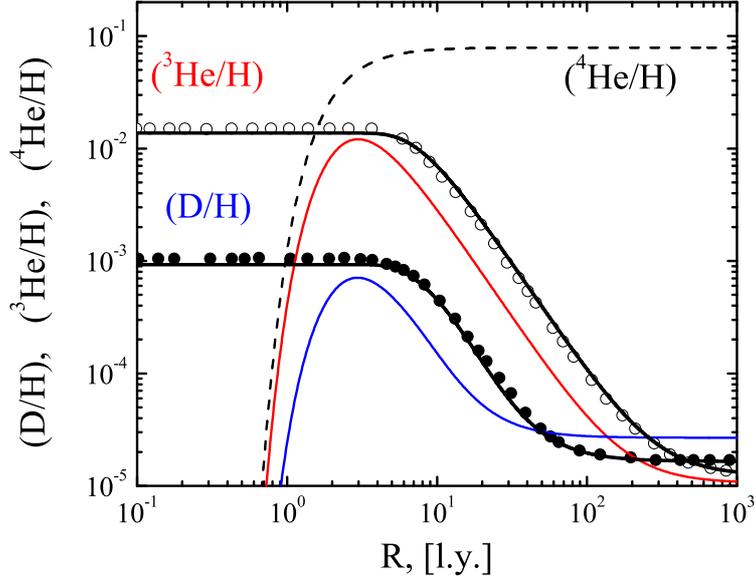} %157
 \caption{~~~($^4$He/H), ($^3$He/H) and (D/H) abundances versus distance to the galactic center of NGC 4151. The circles indicate
the $^3$He/H and D/H curves from Boyd et al. (1989). The solid curves represent our calculations with the data from Boyd
et al. (1989). The dashed curves represent our complete calculations.}
\label{boyd}
\end{figure*}

Previously, the influence of photodisintegration
processes on the abundances of light elements was
considered by Boyd et al. (1989) using the galactic
center of NGC 4151 as an example for the
chosen spectral slope  $\Gamma=2.7$. The gamma-ray
flux with E~$ > 2$~MeV was assumed to be $F=4\times10^{16}$~phot~cm$^2$~s$^{-1}$
 at a distance of 2 lt-days (the
corresponding luminosity is $L_{\gamma}\simeq1.6\times10^{44}$~erg~s$^{-1}$~
in the energy range {1--1000~MeV}). On the whole, our
calculations reproduce the results of this paper, but
there are also differences (see Fig.~\ref{boyd}). In particular,
to estimate the reaction rates, the authors used
reaction cross sections (normalized to the gamma-ray
flux) that turned out to be slightly overestimated.
Table \ref{table3} compares the data from Boyd et al. (1989)
with the values obtained by our fitting of the new data.
The difference between the cross sections for some
reactions reaches 200\%.

Using the same reaction cross sections and initial
element mass fractions as those taken by Boyd
et al. (1989), we reproduced the results of this paper.
However, in the region r$ < $10 lt-years, where the
authors disregarded the change in the abundance of
$^4$He, the results disagree. According to our calculations,
as the galactic center is approached, the
amount of D and $^3$He does not reach a constant level
but reaches its maximum value at $r\sim4-7$  light years.
Subsequently, as the galactic center is approached,
the element abundances will not remain
constant, because the photodisintegration of all elements
begins to dominate as the flux increases.
Accordingly, the change in the abundance of $^4$He
should be taken into account to correctly estimate
the influence of gamma-ray emission on the isotopic
composition of the medium in the immediate neighborhood
of the AGN.

The depletion of $^4$He was taken into account by
Balbes et al. (1996), who also considered the influence
of photonuclear reactions on the primordial
composition of the medium in the Universe. Figure~\ref{balbes}
compares the results of our calculations with
those from Balbes et al. (1996). Our calculations reproduce
the results of this paper. Nevertheless, these
authors did not consider the dependence of the D
and $^3$He production on the spectral parameters of the
emission $L_{\gamma}$ and $\Gamma$ by assuming that the cross sections
for spectra of different hardness increased by
no more than a factor of 4, which actually is not
the case. When varying the spectral hardness, the
reaction cross sections change by one or two orders
of magnitude.

\begin{figure*}
%fig7.
\centering
\includegraphics[width=100mm,clip]{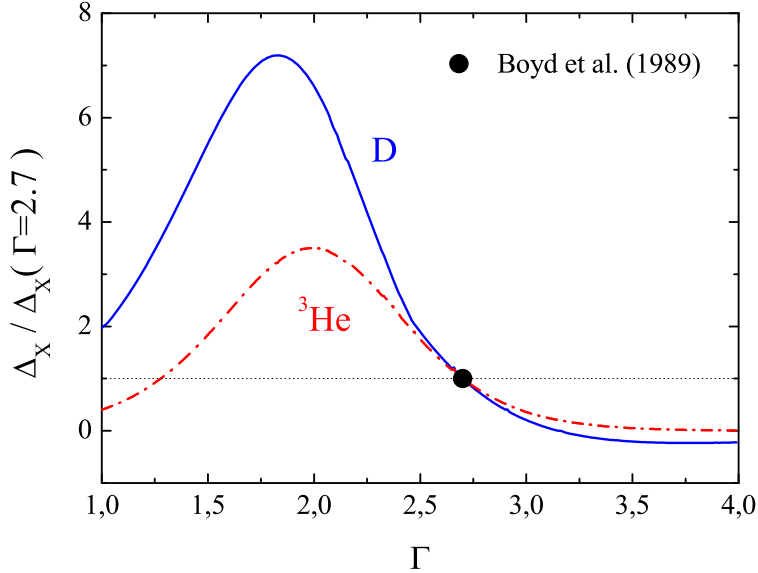} %157
 \caption{Relative amount of produced D and $^3$He through photonuclear reactions in the zone of influence of the galactic center
of NGC 4151 for 1 Gyr versus spectral hardness $\Gamma$. The values are normalized to the amount of D and $^3$He produced when the
medium is exposed to emission with a spectrum with $\Gamma$ = 2.7 (this value was considered by Boyd et al. (1989)). The maximum
amount of D is produced at $\Gamma$ = 1.83 and exceeds the value from Boyd et al. (1989) by more than a factor of 7. The maximum
amount of $^3$He corresponds to $\Gamma$ = 2.0.}
\label{ratioD_he3}
\end{figure*}

In contrast to Boyd~et~al.~(1989) and Balbes~et~al.~(1996),
 we considered the influence of spectral
hardness  $\Gamma$ on the produced amount of D and $^3$He.
We found that the amount of D and $^3$He produced
in the zone of influence of a quasar for spectra of
different values of the spectral index could exceed
the value found by Boyd et al. (1989) by a factor
of 7 for D and a factor of 3 for $^3$He. We retained the
quasar luminosity $L_{\gamma}=1.6\times10^{44}$~erg~s$^{-1}$, changing
only the distribution of gamma-ray photons over the
spectrum (various spectral indices). The derived
dependencies are shown in Fig.~\ref{ratioD_he3}. Thus, the effect
studied by Boyd et al. (1989) using the galactic center
of NGC 4151 as an example may turn out to be
negligible if soft spectra with $\Gamma\ge3$ are considered
or be more significant for harder spectra with $\Gamma\simeq2$
relative to $\Gamma=2.7$ used by Boyd et al. (1989).

\begin{table}
\caption{Comparison of the reaction cross sections.}
\centering
\label{table3}
\vspace{5mm}\begin{tabular}{l|c|c} \hline\hline
   & \multicolumn{2}{|c|}{  $<\sigma>$,~mbarn }\\ %&        <$\sigma$>,~mbarn  \\
   %& \hline
 ~~Reaction     &  ~~~Boyd et al. (1989)~~~       &      ~~~~our calculation~~~~~~            \\
\hline
${\rm \;\,D}(\gamma,n)p            $   & 1.62  &  1.33 \\[5pt]

$^3{\rm He}(\gamma,p){\rm D}       $   & 0.103 &  0.073  \\
$^3{\rm He}(\gamma,n)2p            $   & 0.066 &  0.050  \\[5pt]

$^4{\rm He}(\gamma,p){\rm T}       $   & 0.017 & 0.015 \\

$^4{\rm He}(\gamma,n)^3{\rm He}    $   & 0.017 & 0.015 \\

$^4{\rm He}(\gamma,{\rm D}){\rm D} $   & 5.5$\times10^{-5}$ & 2.62$\times10^{-5}$ \\

$^4{\rm He}(\gamma,pn){\rm D}      $   & 0.0013 & 0.00124 \\

\hline
\end{tabular}
\end{table}

New observations of quasars and blazars (see,
e.g., Arshakian 2011) give an upper limit on the
blazar luminosity in the energy range 1--1000 MeV
of 10$^{49}$~erg~s$^{-1}$. This increases the effective zone
of influence by more than three orders of magnitude
(provided that the activity time of the nucleus is the
same). Note, however, that the emission for blazars
is localized in a narrow region corresponding to the
jet and only in this region is a high value of the characteristic
distance of the influence of emission on the
medium possible. According to our calculations, it is
$\sim$1~kpc for D and $\sim$10~kpc for $^3$He. In the case where
the isotropic source emission model can be applied,
the gamma-ray luminosity of a quasar typically does
not exceed $10^{47}$~erg~s$^{-1}$, i.e., the optical Eddington
luminosity of a quasar with a mass of $10^9~$M$_{sun}$. In
this case, the distance at which the D and $^3$He abundances
change locally are, respectively, $\sim$100~pc and
1~kpc for $^3$He. The derived radii of influence turned
out to be comparable to the characteristic distances
at which the structure of the quasar itself (the source
being non-point like, the accretion disk, and the jet)
should be taken into account. A detailed account	
of the object’s structure can cause the effects to increase.
In the process of its accretion onto an active
nucleus through an accretion disk, being in the immediate
vicinity of the active nucleus, the matter can
be irradiated by considerably larger gamma-ray fluxes
and, accordingly, can change its isotopic composition
in a shorter time. The D- and $^3$He-overrich matter
ejected into the interstellar and intergalactic medium
can cause the isotopic composition of the medium at
great distances from the AGN to change.	
	
\begin{figure*}
%fig7.
\centering
\includegraphics[width=100mm,clip]{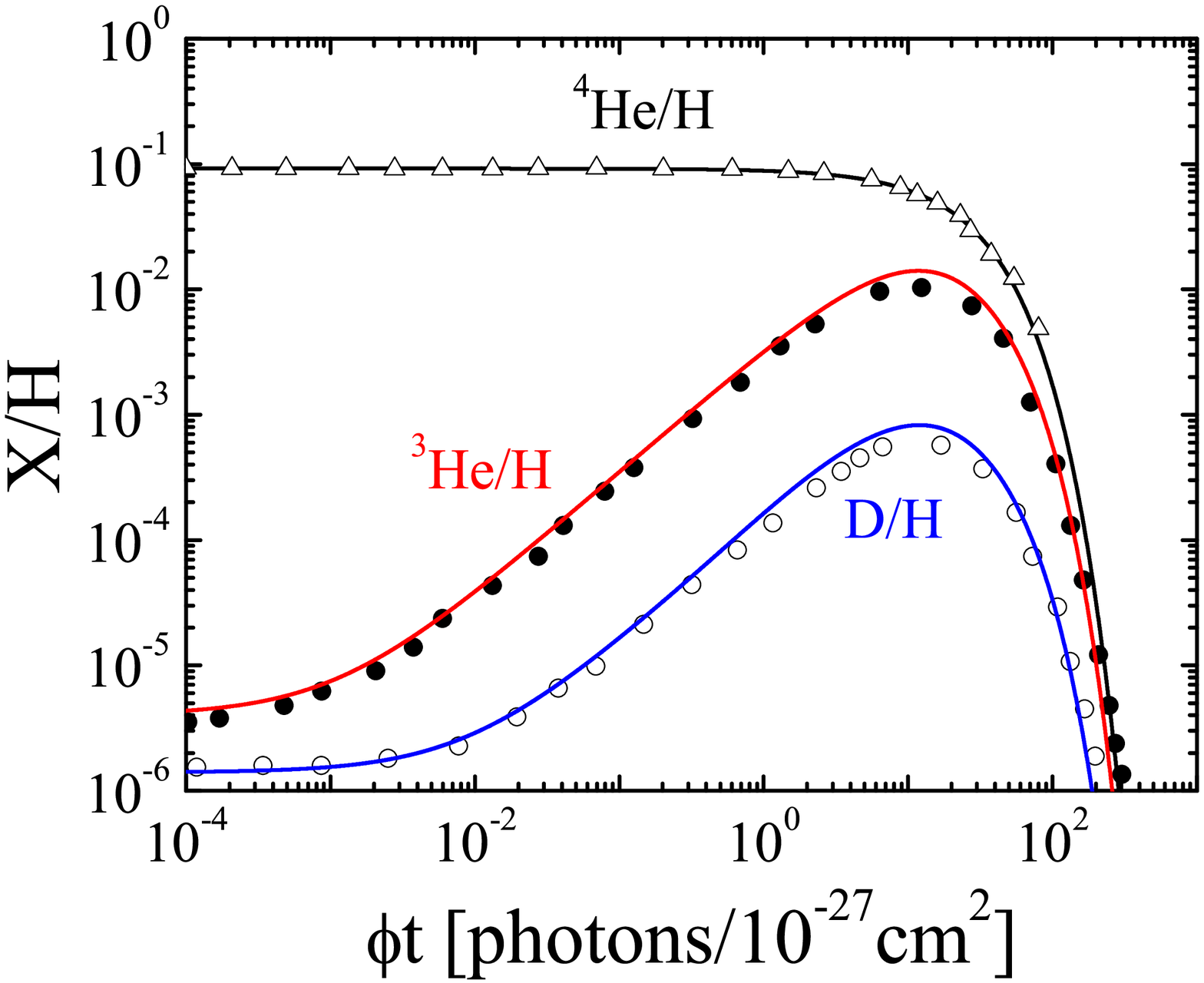} %157
 \caption{($^4$He/H), ($^3$He/H) and (D/H) abundances versus exposure parameter $x=L_{\gamma}t/4\pi r^2$. The dots indicate the $^4$He/H,
$^3$He/H, and D/H curves from Balbes et al. (1996) for $\eta_{10}=100$ (Fig. 1). The solid curves represent our calculations.}
\label{balbes}
\end{figure*}
	
As an example, the application of the examined
mechanism of the change in the isotopic composition
of the medium to our Galaxy can be considered for
the so-called `$^3$He Problem'. Thus, the absence of
a $^3$He/H gradient (Bania et al. 2007) in our Galaxy
allows a limit to be placed on the activity time of the
Galactic nucleus in the past. The hyperfine transition
line of $^3$He+ (8.665 GHz) observed in HII regions
and planetary nebulae is used to determine the
amount of $^3$He in the interstellar medium. The $^3$He/H
ratio is determined via the $^3$He+/H+ ratio. The
$^3$He abundance at Galactocentric distances from 0 to
16 kpc turns out to be at the same level, $^3$He/H=$1.79\pm0.65\times10^{-5}$.
 At a smaller Galactocentric distance,
$^3$He was observed only in the HII region of
the source Sgr B2 in 1990 (Balser 1994): He$^3$/H=2.49$\times10^{-5}$,
 the distance to Sgr B2 was found to be
r=0.09$\pm0.03$~kpc (Reid et al. 2009).

\begin{figure*}
\centering
\includegraphics[width=100mm,clip]{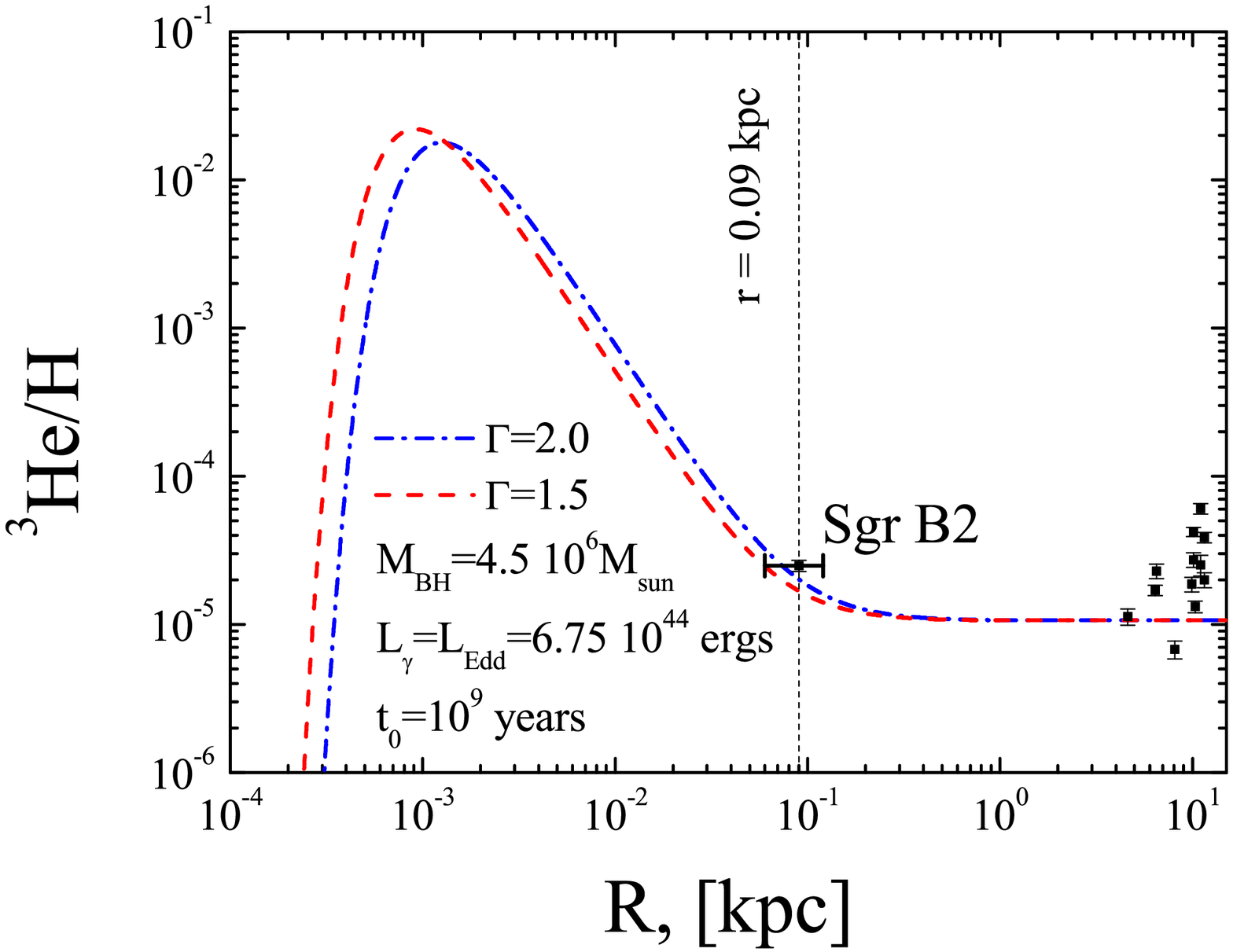} %157
 \caption{~~The dots indicate the experimental data on the $^3$He abundance in the Galaxy as a function of the Galactocentric
distance. The nearest measurement was made in the HII region of Sgr B2. The dashed curve indicates the theoretical calculation
of the $^3$He abundance in a region exposed to gamma-ray emission from the galactic center at the Eddington
luminosity limit for $10^9$~yr. The 3He abundance curve for spectra of different hardness is within the measurement accuracy
of the distance to Sgr B2.}
\label{fig13}
\end{figure*}

Assuming that the increase in $^3$He/H relative to
its primordial value ($^3$He/H)$_{pr}$ was caused only by
photonuclear reactions, we can estimate the activity
time of the Galactic nucleus in the past. The
mass of the supermassive black hole at the center
of our Galaxy, $\mbox{M$_{BH}=4.5\pm0.4\times10^6~$M$_{sun}$}$ (Ghez
et al. 2008), turns out to be so small that to produce
the observed amount of $^3$He at a distance of 100 pc,
the medium should be exposed at the Eddington luminosity
limit of the Galactic nucleus L$_{Edd}=6.75\times10^{44}$~erg~s$^{-1}$
for t$_0\le1.7\pm0.9\times10^{9}$~yr (see Fig. \ref{fig13}),
which exceeds considerably any reasonable estimates
of the lifetime for the nuclei. Some indirect data on
the deuterium abundance point to the existence of a
gradient with Galactocentric distance (Lubowich and
Pasachoff 2009). This can be due to both deuterium
astration in stars and complex chemistry of the interstellar
medium.

\section{CONCLUSIONS}
\label{CONCLUSIONS}
\noindent
We considered the influence of photonuclear reactions
on the change in the composition of the interstellar
medium. Directional exposure of the medium
to gamma-ray emission was shown to cause the relative
D and $^3$He abundances in the medium to increase.

We calculated the photonuclear reaction rates as a
function of the spectral hardness of the emission. We
fitted the photodisintegration reaction cross sections
by analytical formulas using which the reaction rates
can also be expressed analytically in terms of Euler’s
beta functions.

In comparison with previous works (Boyd et al.
1989; Balbes et al. 1996; and others, L~=~10$^{44}$~erg~s$^{-1}$, $\Gamma$~=~2.7)
, we considered power-law spectra with
various values of $\Gamma$ = 1$\div$4 and took into account
the high gamma-ray luminosities of quasars, L$\simeq10^{47}$~erg~s$^{-1}$.
 Allowance for the spectral hardness in the model
 under consideration leads to a change in the final mass
  fractions of the elements by a factor of 3–7.

The global effect from the production of D and $^3$He
through the quasar emission in the Universe, on the
whole, turns out to be insignificant. However, this is
not the case if the changes in the composition of the
medium are considered in a local region. The ejected
matter with a changed isotopic composition can condense
into clouds of interstellar and intergalactic gas,
thereby changing the composition of such clouds. A
further development of the model requires a detailed
analysis of the structure of AGNs, i.e., allowance for
the extent of the jets, the finite sizes of the accretion
disk, and the possibility of the matter processed in the
disk being ejected into the interstellar medium, which
can cause the isotopic composition of the medium at
great distances to change.

Within the model under consideration, the absence
of a 3He/H gradient in the Galaxy constrains
the activity time of the Galactic center at its Eddington
luminosity limit (L$=6.75\times10^{44}$~erg~s$^{-1}$)
$\mbox{$t_0\le1.7\pm0.9\times10^{9}$ yr.}$

\section{ACKNOWLEDGMENTS}
\label{ACKNOWLEDGMENTS}
\noindent
This study was supported by the Ministry of Education
and Science of the Russian Federation (contract
no. 11.G34.31.0001) and the Russian Foundation
for Basic Research (project no. 11-02-01018a).

%\pagebreak
%****************************************************************

%\begin{thebibliography}{99}

%\reference{A. A. Abdo, M. Ackermann,  M. Ajello \etal,
%\apj\ {\bf710}, 1271, (2010).}

%\reference{C.J. Akerman, S.L. Ellison, M. Pettini \etal}, \aap\ {\bf 440}, 499 (2005).

%\end{thebibliography}

%xxxxxxxxxxxxxxxxxxxxxxxxxxxxxxxxxxxxxxxxxxxxxxxxxxxxxxxxxxxxx
%---------------------------------------------------------------
\end{document}